\documentclass[aps]{revtex4}
 \usepackage{amssymb} \usepackage{graphicx}

\begin{document}
 \title{Transgression field theory at the interface of topological insulators}
\author{\"{O}zg\"{u}r A\c{c}{\i}k}
\email{ozacik@science.ankara.edu.tr}
\address{Department of Physics,
Ankara University, Faculty of Sciences, 06100, Tando\u gan-Ankara,
Turkey\\}
 \author{\"Umit Ertem}
 \email{umitertemm@gmail.com}
\address{Department of Physics,
Ankara University, Faculty of Sciences, 06100, Tando\u gan-Ankara,
Turkey\\}

\begin{abstract}
Topological phases of matter can be classified by using Clifford algebras through Bott periodicity. We consider effective topological field theories of quantum Hall
systems and topological insulators that are
Chern-Simons and BF field theories. The edge states of these systems
are related to the gauge invariance of the effective actions. For
the edge states at the interface of two topological insulators,
transgression field theory is proposed as a gauge invariant
effective action. Transgression actions of Chern-Simons theories for
(2+1)D and (4+1)D and BF theories for (3+1)D are constructed. By
using transgression actions, the edge states are written in terms of
the bulk connections of effective Chern-Simons and BF theories.
\end{abstract}

\maketitle

\section{Introduction}

Topological insulators are new phases of matter that generalize the
quantum Hall (QH) systems to different symmetries and dimensions.
One of the characteristic properties of topological insulators for
their classification is time reversal (TR) symmetry and there are
two main classes which are $\mathbb{Z}_2$ insulators that have TR
symmetry and $\mathbb{Z}$ insulators that have no TR symmetry \cite{Hasan Kane,Qi Zhang}. The periodic table of topological insulators and superconductors can be obtained by using the Bott perodicity relation of Clifford algebras in both complex and real cases \cite{Stone Chiu Roy}. Symmetries of corresponding Hamiltonians of topological insulators and superconductors generate the Clifford algebras in relevant dimensions and the relation between symmetric spaces and Clifford algebras gives the topological classes in the periodic table through K-theory methods \cite{Kitaev}.

Topological insulators can also be described effectively by topological field theories. Effective theory of QH systems in two space dimensions ((2+1)D) is
described by Chern-Simons (CS) topological field theory (TFT)
\cite{Zhang Hansson Kivelson,Wen1,Wen2,Fradkin}. Since QH states are
bulk insulating and edge conducting systems and CS theory in (2+1)D
is not invariant under TR symmetry, they correspond to a special case of TR breaking
topological insulators. However, in (4+1)D CS theory is TR invariant
and generalizations of QH states to (4+1)D are TR invariant
topological insulators \cite{Kane Mele,Bernevig
Hughes Zhang,Moore}. Despite the fact that the CS theory is not
defined in odd space dimensions, TFT of TR invariant
topological insulators can be written in (3+1)D by the procedure of
dimensional reduction in terms of axion electrodynamics and the topological theta-term \cite{Qi
Hughes Zhang,Wang Qi Zhang, Huerta Zanelli}. On the other hand, there is also BF
field theory description of topological insulators in (3+1)D
\cite{Cho Moore}. This approach is more convenient for deriving the
edge states from the boundary analysis.

In even space dimensions, effective bulk theory of QH states and
topological insulators are written as abelian CS theories in terms
of a $U(1)$ connection $a$. For manifolds with boundary, CS theory
is not gauge invariant and it transforms under gauge transformations
of the connection $a$ up to a boundary term. To construct the gauge
invariance, one defines a scalar field that lives on the boundary
and add a term to the action that compensates the term spoiling the
gauge invariance. In this way, the edge states are described by the
extra scalar field which do not effect the bulk term
\cite{Wen1,Wen2}. In (3+1)D, BF theory is also not gauge invariant
and similar procedure can be applied to define the edge states of
topological insulators \cite{Cho Moore}. To describe the edge
dynamics completely with the scalar field, one must also add a
kinetic term for the scalar field to the action.

On the other hand, problem of obtaining a gauge invariant action for
CS theories can also have a solution without defining extra fields.
Two CS theories which are defined in two different manifolds and
interact only at their common boundary can be described by a
transgression field theory \cite{Izaurieta Rodriguez Salgado,Mora
Olea Troncoso Zanelli}. A transgression form is defined as whose
exterior derivative is difference of two Chern classes and it can be
written as a sum of two CS theories and a boundary term for two
different bulk connections \cite{Izaurieta Rodriguez Salgado,Mora
Olea Troncoso Zanelli,Nakahara,Bertlmann}. The two CS theories are
defined for two different connections $a$ and $\bar{a}$. By writing
as an action principle for two interacting CS field theories, the
whole transgression action is fully gauge invariant \cite{Borowiec
Ferraris Francaviglia,Borowiec Fatibene Ferraris Francaviglia}. The
boundary term that supply the gauge invariance comes automatically
from the definition of transgression form and the edge dynamics at
the common boundary for two interacting systems can be described by
this boundary term. This construction can be extended to the case of
BF theories and a transgression form for two BF theories interacting
at their common boundary can be written in a similar way. In fact,
this approach may be broadened to more than two TFTs that interact
between themselves at their boundaries \cite{Willison Zanelli}.

In the following sections, we consider two topological insulators described by
two different TFTs and interact at their common boundary in various
dimensions. By using transgression field theory approach, we
investigate the edge excitations from the gauge invariance of the
total action of the system. In (2+1)D, constructing transgression
field theory of two QH states described by abelian CS theories give
the result that the edge excitations are determined by the mean
value of the bulk connections of the two CS theories. In (4+1)D, the
same result is obtained for two TR invariant topological
insulators which have a common interface. In (3+1)D, BF field theory approach of topological
insulators is considered. Up to our knowledge, a transgression
action for two interacting BF theories is absent in the literature.
Therefore we construct a gauge invariant transgression term for two
interacting BF theories under a constraint for gauge transformations
at the boundary. By using this transgression term, we obtain the
result that the edge excitations for two interacting BF theories of
topological insulators with a common interface is determined by the mean value of the bulk
connections with a sign difference from the result obtained for CS
theories. As a result, edge excitations at the interface of two
topological insulators can be described only by bulk connections
without using arbitrary scalar fields.

In the hydrodynamical approach of the edge states for quantum Hall systems \cite{Fradkin}, the edge states are described by fictitious scalar fields that are defined at the boundary of the system. However, in this paper, we find that by using the effective topological field theory approach of quantum Hall systems, the edge states can be obtained from the bulk connections of the theory without needing any additional scalar fields at the boundary. This is done by finding the transgression field theories at the boundary for the bulk topological field theories. This gives a new way to describe the edge states and a simplification for the theory by eliminating the fictitious fields. It also implies that the bulk/edge correspondence through transgression actions for quantum Hall systems and topological insulators gives another manifestation of the holography principle in those systems by only using the bulk properties with no need for defining extra edge degrees of freedom. In that way, the approach used in the paper gives a unified conceptual framework to describe different kinds of topological phases through transgression field theories. Those advances in the theoretical framework of topological states of matter are the main motivations of the paper.

\section{Topological Band Theory and Classification Through Clifford Algebras}

The band structure of free-fermion Hamiltonians can be classified with respect to the symmetries they satisfy. There are ten symmetry classes of free-fermion Hamiltonians with possible anti-unitary symmetries \cite{Ryu Schnyder Furusaki Ludwig}. This tenfold way is originated from the periodicity properties of Clifford algebras. The anti-unitary symmetries such as TR, charge-conjugation and chiral symmetries constitute the generators of the Clifford algebra of the relevant dimension and the following two-fold and eight-fold periodicities of complex and real Clifford algebras respectively \cite{Lawson Michelson}
\begin{eqnarray}
\mathbb{C}l_{n+2}&\cong&\mathbb{C}l_n\otimes\mathbb{C}l_2\nonumber\\
Cl_{n+8,0}&\cong&Cl_{n,0}\otimes Cl_{8,0}\nonumber\\
Cl_{0,n+8}&\cong&Cl_{0,n}\otimes Cl_{0,8}\nonumber
\end{eqnarray}
give the two complex and eight real symmetry classes of topological insulators and superconductors. Here $Cl_{p,q}$ denotes the $p+q$ dimensional real Clifford algebra with $p$ positive and $q$ negative generators and $\mathbb{C}l_p$ denotes the $p$ dimensional complex Clifford algebra. Each of the ten Cartan symmetric spaces correspond to a symmetry class through the relation between representations of Clifford algebras and symmetric spaces \cite{Stone Chiu Roy}. These symmetric spaces play the role of classifying spaces for the symmetry classes and from the homotopy groups and K-theory groups of those classifying spaces, the free-fermion Hamiltonians that have topological phases in relevant dimensions and symmetry classes can be obtained. This correspond to the periodic table of topological insulators and superconductors \cite{Kitaev}. The periodicity relations of Clifford algebras that are given above induce the Bott periodicity of reduced complex and real K-theory groups of $d$-dimensional sphere $S^d$ as follows
\begin{eqnarray}
\widetilde{K}_{\mathbb{C}}(S^{d+2})&=&\widetilde{K}_{\mathbb{C}}(S^{d})\nonumber\\
\widetilde{K}_{\mathbb{R}}(S^{d+8})&=&\widetilde{K}_{\mathbb{R}}(S^{d})\nonumber
\end{eqnarray}
where $\widetilde{K}_{\mathbb{C}}$ denotes the reduced complex K-group and $\widetilde{K}_{\mathbb{R}}$ is the reduced real K-group. This gives way to the two-fold and eight-fold periodicities in complex and real classes in the periodic table \cite{Kitaev}.

The mentioned K-groups can correspond to three different possibilities; the trivial case $0$, integer case $\mathbb{Z}$ and $\mathbb{Z}_2$. The topological phases in $\mathbb{Z}$ classes are called Chern insulators since the topological invariants that characterize them are Chern and winding numbers and the topological phases in $\mathbb{Z}_2$ classes are called $\mathbb{Z}_2$ insulators.

\section{Topological Field Theory and Transgression Forms}

The classification of topological phases of matter can be described by effective field theories in the low energy limit. The effective field theories of topological insulators correspond to topological field theories and they can be written in terms of topological terms in the action and relevant topological invariants.

\subsection{(2+1)D and (4+1)D}

In (2+1)D, bulk effective action of QH systems is given by an
abelian CS theory in terms of a $U(1)$ connection $a$;
\begin{equation}
S_0=\frac{C_1}{4\pi}\int_M a\wedge da.
\end{equation}
Here $\wedge$ is the wedge product and $d$ is the exterior derivative operator. This action is not gauge invariant under the transformations
$a\rightarrow a+d\gamma$ for a manifold $M$ with boundary $\partial
M$ where $\gamma$ is a function. The constant coefficient $C_1$ represents the quantized Hall
conductivity and given by the first Chern number of the Berry
curvature of the system \cite{Thouless Kohmoto Nightingale den
Nijs};
\begin{equation}
C_1=\frac{1}{2\pi}\int d^2k\epsilon^{ab}f_{ab}
\end{equation}
where the integral is taken over the momentum space, $\epsilon_{ab}$
is the antisymmetric Levi-Civita symbol of valence two and
$f_{ab}$'s are the components of Berry curvature. $a,b$ takes values
$0,1,2$ and the summation convention is assumed. The Chern number
gives the topological phase of the system and it takes integer
values. However, this coefficient does not affect the gauge
invariance properties of the action. The edge dynamics of QH states
stems from restoring the gauge invariance of the action by adding a
boundary term;
\begin{equation}
S=\frac{C_1}{4\pi}\int_M a\wedge da+\frac{C_1}{4\pi}\int_{\partial
M}d\phi\wedge a
\end{equation}
where $\phi$ is a scalar field that transforms as $\phi\rightarrow
\phi+\gamma$ and it is defined from the gauge condition at the
boundary $a_0=0$ and $a_i=\partial_i\phi$ for $i=1,2$
\cite{Wen1,Wen2} and it should be noted that, since $\phi$ is a
boundary field, its exterior derivative has no bulk part; that is
$d\phi=d|_{\partial M}\phi$. In fact, the Hamiltonian of the
boundary term is zero and the construction of the edge dynamics
needs an extra kinetic term which is $\int_{\partial
M}D\phi\wedge*D\phi$ where $D\phi=d\phi-a$ and $*$ is the Hodge star
operation on differential forms over $\partial M$.

The edge excitation $\phi$ which is determined from the boundary
terms of the gauge invariant action is related to the electron
operator $\Psi$ at the edge as follows \cite{Wen1,Wen2,Fradkin}
\begin{equation}
\Psi\propto\exp{\left(i\frac{1}{C_1}\phi\right)}
\end{equation}
and $\Psi$ satisfies the fermion statistics for $C_1^{-1}$ is an odd
integer which means that we will consider the cases of $C_1=1$. In hydrodynamical approach of the edge excitations, the
edge waves are described by the density
\begin{eqnarray}
\rho=\frac{1}{2\pi}\partial_x\phi\nonumber
\end{eqnarray}
and it satisfies the wave
equation
\begin{eqnarray}
\partial_t\rho-v\partial_x\rho=0.\nonumber
\end{eqnarray}
Here, $t$ denotes the time
coordinate, $x$ is the coordinate that is parallel to the boundary
and $v$ is the group velocity of the edge excitation. The density
$\rho$ is defined as $\rho(x)=nh(x)$ where $h(x)$ is the
displacement of the edge and $n$ is the electron density in the
bulk. The quantization of the Hamiltonian of the edge waves gives
the relation (4) for the fermionic operator that carries the charge
$e$. The sign of the velocity of the edge excitations is determined
by the sign of the Chern number \cite{Wen1,Wen2}.

Let us consider two QH states defined for the connections $a$ and
$\bar{a}$ living on the cobordant manifolds $M$ and $\bar{M}$
respectively, namely $\partial M=\partial \bar{M}$, with Chern
numbers $C_1$ and $C_1'$ and have an interface along their common boundary.
The full action without kinetic terms is written as
\begin{eqnarray}
S&=&\frac{C_1}{4\pi}\int_M a\wedge da+\frac{C_1}{4\pi}\int_{\partial
M}d\phi\wedge a\nonumber\\
&+&\frac{C_1'}{4\pi}\int_{\bar{M}} \bar{a}\wedge
d\bar{a}+\frac{C_1'}{4\pi}\int_{\partial M}d\bar{\phi}\wedge \bar{a}
\end{eqnarray}
A special choice for the Chern numbers of two QH states can be
$C_1'=-C_1$. Then, the action of the system is as follows
\begin{equation}
S=\int_M a\wedge da+\int_{\partial M}d\phi\wedge a-\int_{\bar{M}}
\bar{a}\wedge d\bar{a}-\int_{\partial M}d\bar{\phi}\wedge \bar{a}
\end{equation}
where the overall constant $C_1/4\pi$ is not written for simplicity,
since it does not affect the gauge invariance of the action, but it
should be kept in mind that $C_1$ has to be unity ($C_1=1$) for
statistical reasons. This choice of Chern numbers will also be
relevant for the following discussions of (4+1)D and (3+1)D cases.
By the way, since we have a common boundary for two different QH
systems, the edge dynamics must be same and the scalar fields should
be taken as $\phi=\bar{\phi}$.

On the other hand, a gauge invariant action for two CS theories
${\mathcal{C}}_{2n+1}(a)$ and ${\mathcal{C}}_{2n+1}(\bar{a})$ interacting
along their common boundary can be written in terms of transgression
form without using extra scalar fields \cite{Izaurieta Rodriguez
Salgado,Mora Olea Troncoso Zanelli};
\begin{equation}
S_T=\int_M
{\mathcal{C}}_{2n+1}(a)-\int_{\bar{M}}{\mathcal{C}}_{2n+1}(\bar{a})-\int_{\partial
M}B_{2n}(a,\bar{a})
\end{equation}
where
\begin{equation}
{\mathcal{C}}_{2n+1}(a)=(n+1)\int_0^1dt\langle a(tda+t^2a^2)^n\rangle
\end{equation}
is the CS (2n+1)-form and
\begin{equation}
B_{2n}=-n(n+1)\int_0^1sds\int_0^1dt\langle a_t (a-\bar{a})
f_{st}^{n-1}\rangle
\end{equation}
is the boundary term. Here, $\langle...\rangle$ is the symmetric invariant trace for the
Lie algebra and $a_t=ta+(1-t)\bar{a}$ is the composed connection of $a$ and $\bar{a}$ with the curvature $f_t=da_t+a_t\wedge a_t$.
The definition of $f_{st}$ is $f_{st}=sf_t+s(s-1)a_t\wedge a_t$ and $f_t^n=f_t\wedge...\wedge f_t$
($n$ times). For (2+1)D abelian CS theories the gauge invariant
transgression action is written as follows (by taking $n=1$ in (7))
\begin{equation}
S_T=\int_M a\wedge da-\int_{\bar{M}}\bar{a}\wedge
d\bar{a}-\int_{\partial M}a\wedge\bar{a}.
\end{equation}
By comparing (10) and (6) one obtains that the edge dynamics of two
QH states at their common boundary can be written from
the equality
\begin{eqnarray}
\frac{a+\bar{a}}{2}=d\phi\nonumber
\end{eqnarray}
which is relevant only at the
boundary. Hence, we found that to construct the edge dynamics of the system, there is no need to
define a gauge choice at the boundary and the gauge invariance of
the transgression action gives the form of the scalar field in terms
of the bulk connections. So, transgression field theory serves a
natural gauge invariant effective theory for two QH
systems with a common interface using only bulk connections.

The first TR invariant topological insulator state was introduced
for (4+1)D \cite{Zhang Hu,Bernevig Chern Hu Toumbas Zhang} and TR
invariant topological insulators for other dimensions can be
constructed from this root state. The effective theory of this state
is given by the abelian CS theory in (4+1)D;
\begin{equation}
S_0=\frac{C_2}{24\pi^2}\int_M a\wedge da\wedge da.
\end{equation}
In this case $C_2$ is given by the second Chern number of the Berry
curvature;
\begin{equation}
C_2=\frac{1}{32\pi^2}\int d^4k\epsilon^{abcd}tr[f_{ab}f_{cd}]
\end{equation}
where $a,b,c,d$ takes values $0,1,2,3,4$. The gauge invariance
requirements again modify the action with a boundary term for a
scalar field $\phi$;
\begin{equation}
S=\frac{C_2}{24\pi^2}\int_M a\wedge da\wedge
da+\frac{C_2}{24\pi^2}\int_{\partial M}d\phi\wedge a\wedge da
\end{equation}
and a kinetic term for the dynamics of the boundary scalar field. By
considering two TR invariant topological insulator states
with a common interface as in the case of (2+1)D with
Chern numbers $C_2'=-C_2$, one can write the total action for two CS
theories as
\begin{eqnarray}
S&=&\int_M a\wedge da\wedge da+\int_{\partial M}d\phi\wedge a\wedge
da\nonumber\\
&-&\int_{\bar{M}} \bar{a}\wedge d\bar{a}\wedge
d\bar{a}-\int_{\partial M}d\bar{\phi}\wedge \bar{a}\wedge d\bar{a}.
\end{eqnarray}
However, we can use the transgression field theory for this system
and write the following action from (7) (by taking $n=2$)
\begin{equation}
S_T=\int_M a\wedge da\wedge da-\int_{\bar{M}}\bar{a}\wedge
d\bar{a}\wedge d\bar{a}-\int_{\partial M} a\wedge
\bar{a}\wedge(da+d\bar{a}).
\end{equation}
The comparison of two CS theories written as in (14) for
$\phi=\bar{\phi}$ and the transgression action (15) gives the result
that the scalar field which define the edge dynamics of this system
is given as
\begin{eqnarray}
\frac{a+\bar{a}}{2}=d\phi.\nonumber
\end{eqnarray}
This is exactly the same result
that we found in (2+1)D case and this shows that the transgression field theory
approach for topological insulators with a common interface is consistent for
different dimensions in defining boundary scalar fields from the
bulk connections.

\subsection{BF theory and (3+1)D}

In (3+1)D, TR invariant topological insulators are described by an
effective topological action that is the second Chern class of an
abelian $U(1)$ connection with a coefficient $\theta$ which takes
only two values \cite{Qi Hughes Zhang}. On the other hand, a
different approach to describe the TR invariant topological
insulators in (3+1)D in terms of a topological field theory is also
recently proposed \cite{Cho Moore}. In this case, the topological BF
field theory is the effective action for (3+1)D TR invariant
topological insulators which is given by
\begin{equation}
S_0=\theta\int_M b\wedge f
\end{equation}
where $b$ is a 2-form field and $f=da$ is the curvature of the
abelian $U(1)$ connection $a$. $\theta$ is the $\mathbb{Z}_2$
topological invariant that characterizes the TR invariant
topological insulators in (3+1)D. The $b$ field corresponds to the
response of the system to an applied $\pi$ flux (which corresponds
to the response of the system to an external potential) \cite{Cho
Moore}. However, like in the CS case, this action is also not gauge
invariant and to satisfy the gauge invariance one needs to add
boundary terms to the action. For a boundary scalar field $\phi$,
that gives the edge dynamics of the system, the following action is
gauge invariant under the transformations $b\rightarrow b+d\xi$,
$a\rightarrow a+d\gamma$ and $\phi\rightarrow\phi-\gamma$;
\begin{equation}
S=\theta\int_M b\wedge f-\theta\int_{\partial M}b\wedge
a-\theta\int_{\partial M}b\wedge d\phi.
\end{equation}
If we consider two topological insulators which have a
common boundary with different topological invariants as before, the
total action is written as follows
\begin{eqnarray}
S&=&\int_M b\wedge f-\int_{\partial M}b\wedge a-\int_{\partial
M}b\wedge d\phi\nonumber\\
&-&\int_{\bar{M}} \bar{b}\wedge \bar{f}+\int_{\partial
M}\bar{b}\wedge \bar{a}+\int_{\partial M}\bar{b}\wedge d\bar{\phi}
\end{eqnarray}
and we take $\phi=\bar{\phi}$ for the common boundary because of we have the same boundary for two topological insulators which implies that the boundary degrees of freedom must be equal to each other.

However, unlike the CS case, there is no transgression form defined
in the literature for two BF field theories interacting along their
common boundary. Hence, to do the same boundary analysis as before
we have to define a transgression action for BF theories. In fact,
this can be done in (3+1)D with some constraints on gauge
transformations. Let us consider the following action for two BF
theories with common boundary
\begin{equation}
S_{TBF}=\int_M b\wedge
f-\int_{\bar{M}}\bar{b}\wedge\bar{f}-\frac{1}{2}\int_{\partial
M}(b+\bar{b})\wedge(a-\bar{a}).
\end{equation}
This action is gauge invariant under the transformations
\begin{equation}
b\rightarrow b+d\xi\quad,\quad \bar{b}\rightarrow
\bar{b}+d\bar{\xi}\quad,\quad a\rightarrow a+d\gamma\quad,\quad
\bar{a}\rightarrow \bar{a}+d\bar{\gamma}
\end{equation}
if $\bar{\gamma}=\gamma+c$ and $\bar{\xi}=\xi+d\alpha$, where $\xi$ is a 1-form, $c$ is
a constant and $\alpha$ is a function. This gauge invariance property shows that the action (19) corresponds to a transgression action. Hence, we construct a
transgression action for abelian (3+1)D BF theories for some special
gauge transformations.

Now, we can do the same comparison between two interacting BF
theories (18) and the transgression action (19) and find that the
scalar field that define the dynamics at the boundary is given by
\begin{eqnarray}
-\frac{a+\bar{a}}{2}=d\phi.\nonumber
\end{eqnarray}
So, the transgression field theory is also
relevant for (3+1)D BF theories case and the exterior derivative of
the scalar field is again proportional to the mean value of the bulk
connections as before. This construction strengthens the
transgression field theory approach that we propose for the boundary dynamics of
topological insulators.

\section{Conclusion}

The edge dynamics of topological insulators is a
result of gauge invariance of the bulk topological action. However,
the dynamics is described by a boundary scalar field in the literature. If we have two
topological insulators with an interface at their common boundary, then we
can write a gauge invariant action for this system as a
transgression field theory. By doing this, we show that the dynamics at the
boundary is determined by the mean value of the bulk connections;
\begin{equation}
\pm\frac{a+\bar{a}}{2}=d\phi
\end{equation}
and this is relevant for both (2+1)D and (4+1)D CS theory and (3+1)D
BF theory cases. This is the main result of the paper and this implies that the density of the edge waves are determined
by the bulk connections as follows
\begin{eqnarray}
\rho&=&\frac{1}{2\pi}\sum_i\partial_{x_i}\phi=\pm\frac{1}{4\pi}\sum_i(a_i+\bar{a}_i)
\end{eqnarray}
where $x_i$ are the coordinates of the boundary surface and
$a_i$ and $\bar{a}_i$ are components of the bulk connections. Hence, the number density at the boundary is described by the intrinsic gauge fields $a$ and $\bar{a}$. From (21), one obtains that the boundary scalar field can be written as a sum of the holonomies of the two bulk connections along the boundary
\begin{equation}
\phi=\pm\frac{1}{2}\left(\int_{\partial M}a+\int_{\partial M}\bar{a}\right).
\end{equation}
Hence, the electron operator at the edge is written in terms of holonomies from (4) as
\begin{equation}
\Psi\propto\exp{\left(\pm\frac{i}{2C_1}\left(\int_{\partial M}a+\int_{\partial M}\bar{a}\right)\right)}.
\end{equation}
While the connections are defined in the bulk, holonomies are defined as integrals around the boundary. This is a correspondence between bulk and boundary and another manifestation of the holography principle in topological insulators.

As a result of (21), the electron operator $\Psi$ at the interface
of two topological insulators with opposite Chern numbers is
determined by the bulk connections through the relation (24). In view
of this, the transgression actions are natural candidates for
effective TFTs of two topological insulators with common boundary
and the edge dynamics is determined by the bulk connections of
transgression field theory. In that way, we show that the edge states can be completely determined by the bulk connections without using any arbitrary scalar fields and the transgression field theory approach gives a new unified description for the edge states of different topological phases.


\end{document}